\documentclass[twoside]{article}

\usepackage{marvosym}
\usepackage[preprint]{aistats2027}
\renewcommand{\toptitlebar}{\vskip .25in}
\renewcommand{\bottomtitlebar}{\vskip .1in}
\usepackage[round,authoryear]{natbib}
\usepackage{graphicx}
\usepackage[hyphens]{url}
\usepackage{algorithm}
\usepackage{algpseudocode}
\usepackage{booktabs}
\usepackage{amsfonts}
\usepackage{amsmath}
\usepackage{amssymb}
\usepackage{multirow}
\usepackage{microtype}
\usepackage{rotating}
\usepackage{tabularx}
\usepackage{array}

\begin{document}
\runningtitle{Intervention-Conditioned Human Digital Twin Generator}
\runningauthor{Farzi et al.}

\twocolumn[
\aistatstitle{An Intervention-Conditioned Human Digital Twin Generator for In Silico Clinical Trial Simulation}

\aistatsauthor{
Saeed Farzi$^{1}$,
Sofia Michopoulou$^{2,3}$,
Christopher Kipps$^{2,3}$,
Rahman Attar$^{1}$\textsuperscript{\Letter}
}
\vspace{.2in} % Gap between authors and affiliations

\aistatsaddress{
$^{1}$School of Electronics and Computer Science,
University of Southampton, Southampton, UK\\
\texttt{r.attar@southampton.ac.uk}\\[4pt]
$^{2}$University Hospital Southampton NHS Foundation Trust,
Southampton, UK\\
$^{3}$Faculty of Medicine, University of Southampton,
Southampton, UK
}

% \aistatsauthor{Saeed Farzi$^{1}$, Sofia Michopoulou$^{2,3}$, Christopher Kipps$^{2,3}$, Rahman Attar$^{1}$, \\ and for the Alzheimer's Disease Neuroimaging Initiative\textsuperscript{*}}

% \aistatsauthor{Saeed Farzi$^{1}$, Sofia Michopoulou$^{2,3}$, Christopher Kipps$^{2,3}$, Rahman Attar$^{1}$}

% \aistatsaddress{and for the Alzheimer's Disease Neuroimaging Initiative\textsuperscript{+}\\$^{1}$School of Electronics and Computer Science, University of Southampton, Southampton, UK\\
% $^{2}$University Hospital Southampton NHS Foundation Trust, Southampton, UK\\
% $^{3}$Faculty of Medicine, University of Southampton, Southampton, UK}

]

% \begingroup
% \renewcommand{\thefootnote}{+}
% \footnotetext{Data used in preparation of this article were obtained from the Alzheimer's Disease Neuroimaging Initiative (ADNI) database (\url{https://adni.loni.usc.edu/}). As such, the investigators within ADNI contributed to the design and implementation of ADNI and/or provided data but did not participate in the analysis or writing of this report. A complete listing of ADNI investigators can be found at \url{https://adni.loni.usc.edu/wp-content/uploads/how_to_apply/ADNI_Acknowledgement_List.pdf}.}
% \endgroup

% \footnotetext{+Data used in preparation of this article were obtained from the Alzheimer's Disease Neuroimaging Initiative (ADNI) database (\url{https://adni.loni.usc.edu/}). As such, the investigators within ADNI contributed to the design and implementation of ADNI and/or provided data but did not participate in the analysis or writing of this report. A complete listing of ADNI investigators can be found at \url{https://adni.loni.usc.edu/wp-content/uploads/how_to_apply/ADNI_Acknowledgement_List.pdf}.}
\begin{abstract}
Chronic diseases evolve through long, heterogeneous trajectories shaped by interacting mediators, confounders, and treatment histories. Estimating long-term patient trajectories under alternative interventions is essential for in silico clinical trial design. However, most existing models overlook irregular visit intervals and cross-mediator dependencies, both of which are crucial for reliable long-term longitudinal prediction. Moreover, these models typically focus on outcome prediction rather than generating complete patient-state trajectories. To address these challenges, we propose CHDT, an intervention-conditioned Human Digital Twin generator under standard identification assumptions for multimodal longitudinal trajectory estimation. Its main
empirical advantage arises from explicit modeling of irregular visit intervals, while cross-mediator refinement provides
additional gains in selected intervention-conditioned settings. Across real-world ADNI and simulator-generated ALPACA cohorts, CHDT shows improved long-horizon patient-state fidelity relative to competing baselines. Ablation results indicate that irregular-time conditioning provides the primary empirical gain, and cross-mediator refinement provides additional improvements.

\end{abstract}
%\link{Code  Omitted for blind review}{}
%\link{Datasets   Omitted for blind review}{}
\section{Introduction}

Chronic noncommunicable diseases, including cardiovascular disease, cancer, dementia, diabetes, and chronic respiratory disease, remain a major global health challenge. According to the World Health Organization, chronic diseases caused at least 43 million deaths globally in 2021~\citep{who_ncd_2025}. These conditions often progress over long and heterogeneous trajectories shaped by interacting biological, physiological, environmental, behavioral, and treatment-related factors. Capturing such trajectories is therefore essential for risk stratification, early deterioration detection, treatment-response forecasting, and in silico clinical trial design.

Human Digital Twins (HDTs) offer a promising framework for modeling disease progression, supporting intervention planning, and enabling in silico clinical trial simulation. However, many existing HDT models remain largely predictive or correlational, limiting their ability to simulate how patient trajectories may evolve under alternative treatment policies~\citep{vallee2026prediction,makarov2025large}. Although recurrent neural networks, temporal transformers, and generative sequence models have advanced longitudinal disease modeling and temporal forecasting, they often focus on event forecasting, electronic health record generation, or short-term outcome prediction rather than full patient-state trajectory generation~\citep{xiao2025xtsformer,lu2026lumin,seedat2022continuous}. This limitation is particularly important in chronic diseases such as dementia, where clinical histories are long, sparse, irregularly sampled, and multimodal, and where interventions, elapsed time, and interacting clinical states jointly shape disease progression~\citep{vallee2026prediction,ishwaran2025harnessing,moen2025towards,makarov2025large}.

To address these challenges, we propose CHDT, an intervention-conditioned Human Digital Twin generator for long-term multimodal trajectory estimation and in silico treatment-arm simulation. CHDT generates full future patient-state trajectories under treatment policies from sparse, irregular, multimodal clinical histories and static patient characteristics. It explicitly models irregular visit intervals to support time-aware autoregressive rollouts over unequal follow-up intervals. In addition, CHDT incorporates cross-mediator dependency modeling, allowing interactions among clinical variables to be captured through mediator-specific decomposition without requiring full joint-state modeling.

We evaluate CHDT through a comprehensive experimental framework covering factual trajectory fidelity, intervention-conditioned counterfactual fidelity, counterfactual plausibility, and trajectory generation performance. Experiments are conducted across four complementary scenarios using real-world ADNI data~\citep{petersen2010adni} and synthetic data generated using ALPACA~\citep{brady2026alpaca}. The results suggest that CHDT achieves consistent overall performance, particularly in preserving disease-state fidelity. Ablation analysis indicates that irregular visit interval modeling and cross-mediator dependency learning improve trajectory quality, while plausibility analysis suggests that generated intervention trajectories largely remain within empirically supported treatment regions.

The main contributions of this paper are threefold. First, we formulate Human Digital Twin generation as an intervention-conditioned multimodal trajectory modeling problem, enabling patient-specific rollout of complete future clinical states under specified treatment policies. Second, we introduce a longitudinal factorization that explicitly incorporates irregular visit intervals and cross-mediator dependencies during autoregressive patient-state generation. Third, we develop an evaluation framework based on real-world and simulator-generated cohorts to assess factual trajectory fidelity, intervention-conditioned fidelity, cross-source robustness, and empirical treatment support.

\section{Related Work}
 \paragraph{Human Digital Twins}
\citet{makarov2025large} proposed DT-GPT, a large language model framework for forecasting patient health trajectories from longitudinal medical data. Similarly, \citet{moen2025towards} used transformer-based architectures to model evolving longitudinal health trajectories. In another direction, personalized cardiac digital twins have been employed to simulate patient-specific arrhythmia progression~\citep{rajawat2025cardiac}.
 \paragraph{Causal Treatment Models}
A parallel research direction focuses on causal inference reasoning in longitudinal healthcare modeling. GNET~\citep{li2021g} introduced a recurrent neural network to estimate counterfactual trajectories. More recently, transformer-based causal models, such as GTrans~\citep{hess2024igc}, have extended neural g-computation to capture temporal dynamics. Recurrent Marginal Structural Networks (RMSN) combine recurrent sequence modeling with inverse-probability weighting~\citep{lim2018forecasting}. Counterfactual Recurrent Networks (CRN) extend this direction by learning balanced recurrent representations through adversarial training~\citep{bica2020estimating}. TECDE uses neural controlled differential equations to model counterfactual outcomes in continuous time~\citep{seedat2022continuous}.

\paragraph{Positioning of this Work}
Unlike predictive Human Digital Twin frameworks that primarily model longitudinal forecasting and causal sequence models that typically estimate future treatment outcomes, CHDT targets intervention-conditioned generation of complete multimodal patient-state trajectories. Its contribution lies in combining irregular-time autoregressive generation with a mediator-specific dependency decomposition that captures interactions among generated clinical states without requiring explicit modeling of the full joint mediator distribution.
\section{Methodology}
\subsection{Problem Formulation}
CHDT generation is formulated as a conditional counterfactual trajectory distribution estimation problem as follows.

\paragraph{Integrated Clinical State Representation}
For a patient $p$ observed over $t$ clinical visits, the longitudinal trajectory is denoted by
$
X_{1:t}^{p} = \{X_1^{p}, X_2^{p}, \ldots, X_t^{p}\}
$
where $X_i^{p}$ represents the Integrated Clinical State (ICS) of the patient $p$ at visit $i$. For simplicity, we omit the patient index $p$ and use uppercase letters for random variables and lowercase letters for their realized values.

Guided by causal mediation analysis~\citep{robins1986new}, in which treatment $A$ affects outcome $Y$ through both direct and indirect mediator-mediated dependencies, ICS is represented as:
\begin{equation}
X_i = \langle M_i, A_i, Y_i \rangle,
\label{eq:visit-representation}
\end{equation}
where $M_i$, $A_i$, and $Y_i$ denote the set of multimodal time-varying mediator states, the observed treatment or action, and the clinical outcome of interest at visit $i$, respectively.

The vector $s$ captures time-invariant patient characteristics, including sex, genetic factors, and demographic information. Since real-world clinical data are rarely collected at fixed intervals due to unequal follow-up schedules, missed visits, or delayed assessments, $\Delta t=\{\delta t_1,\delta t_2,\ldots\}$ denotes the sequence of elapsed time intervals between successive visits.
The factual observed history of a patient up to visit $b$ is defined as:
$
F_b = \{X_1, X_2, \ldots, X_b\}
$
while the future counterfactual trajectory from visit $b+1$ to visit $t$ under an intervention policy, $\pi = \{a_{b+1}^{\pi}, a_{b+2}^{\pi}, \ldots, a_t^{\pi}\}$, is denoted by
$
X_{b+1:t}^{\pi} = \{X_{b+1}^{\pi}, X_{b+2}^{\pi}, \ldots, X_t^{\pi}\}.
$

\paragraph{Counterfactual Human Digital Twin Generator}
Rather than estimating only a single future outcome, the objective is to estimate the full counterfactual trajectory distribution. A CHDT generator is therefore defined as follows.
\begin{equation}
f_{\Theta}\left(X_{b+1:t}^{\pi}\mid F_b, \Delta t, s, \pi\right)
\approx
p\left(X_{b+1:t}^{\pi} \mid F_b,\Delta t, s, \pi\right),
\label{eq:digital-twin-generator}
\end{equation}
where $f_{\Theta}(\cdot)$ is the learnable CHDT generator parameterized by $\Theta$. During inference, multiple trajectories are sampled as:
\begin{equation}
\hat{X}_{b+1:t}^{\pi,j}
\sim
f_{\Theta}(\cdot\mid F_b,\Delta t,s,\pi),
\quad j=1,\dots,J.
\label{eq:mc-rollouts}
\end{equation}

Under standard causal assumptions, CHDT estimates intervention-conditioned counterfactual trajectories using an autoregressive process inspired by longitudinal g-computation. These counterfactual interpretations depend on assumptions including consistency, sequential exchangeability, and positivity~\citep{robins1986new}.

\subsection{Proposed Solution}
To estimate $\hat{X}_{b+1:t}^{\pi}$, the model sequentially estimates the next-step outcome distribution together with the corresponding multimodal mediator states during Monte Carlo simulations.
Inspired by~\citet{robins1986new}, the counterfactual outcome distribution is factorized as follows:
\begin{equation}
\small
\begin{aligned}
p\left(Y_t=y\mid F_b,s,\pi, \Delta t\right)=
\\
\int_{m_{b+1:t}}
&p\left(
Y_t=y \mid H_{t-1}^{\pi}, m_t, a_t^{\pi}, \delta t_t
\right)
\\
&
\prod_{k=b+1}^{t}
p\left(
M_k=m_k
\mid
H_{k-1}^{\pi}, a_k^{\pi}, \delta t_k
\right)
\end{aligned}
\label{eq:ctd1}
\end{equation}

where $m_{b+1:t}$ denotes one possible future mediator trajectory and $
H_{k-1}^{\pi}=
\{F_b,
X_{b+1:k-1}^{\pi},
s,
\Delta t_{b+1:k-1}
\}$ defines the autoregressive counterfactual history at future step $k$. In Eq.~\ref{eq:ctd1}, the outcome $Y_t$ and mediator states $M$ can be treated in the same way, without loss of generality~\citep{li2021g}. Exact computation of Eq.~\ref{eq:ctd1} is generally infeasible in practice because it would require summing or integrating over all possible mediator trajectories, whose number grows rapidly with the number of mediators and the prediction horizon. Therefore, CHDT approximates it using a finite number of Monte Carlo simulations.

\subsection{Mediator Dependency Decomposition}
We assume that $M_k$ is represented as a collection of $|M|$ mediators:
$M_k =
\langle
M_{k}^{1}, M_{k}^{2}, \ldots, M_{k}^{|M|}
\rangle$.
Directly modeling the full joint dependency structure among mediators is computationally challenging, especially when the number of mediators $|M|$ is large. To reduce this complexity, the proposed model first estimates mediator-specific prior heads without explicitly conditioning on other mediator states:
\begin{equation}
\tilde{M}_k^{j,\pi}
\sim
p_{\text{prior}}^{j}
\left(
M_k^j
\mid
H_{k-1}^{\pi},a_k^\pi,\delta t_k
\right),
\quad j=1,\dots,|M|.
\label{eq:mediator-prior}
\end{equation}
Then, each mediator-specific state is refined by conditioning on the prior estimates of all other mediators, excluding its own prior:
\begin{equation}
\hat{M}_k^{j,\pi}
\sim
p_{\text{refine}}^{j}
\left(
M_k^j
\mid
H_{k-1}^{\pi},a_k^\pi,\delta t_k,\tilde{M}_k^{-j,\pi}
\right),
\label{eq:mediator-refine}
\end{equation}
where
\[
    \tilde{M}_k^{-j}
    =
    \{\tilde{M}_k^r : r \neq j\}.
    \]
This two-stage process allows CHDT to capture cross-mediator dependencies without directly modeling the full joint mediator distribution. Categorical outputs can use softmax heads, while continuous outputs can use Gaussian or regression heads. Note that this refinement module captures conditional dependence among generated clinical states; it is not interpreted as recovering the underlying structural causal graph.

\subsection{Model Architecture}

\begin{figure*}[t]
    \centering
    \includegraphics[width=0.9\linewidth]{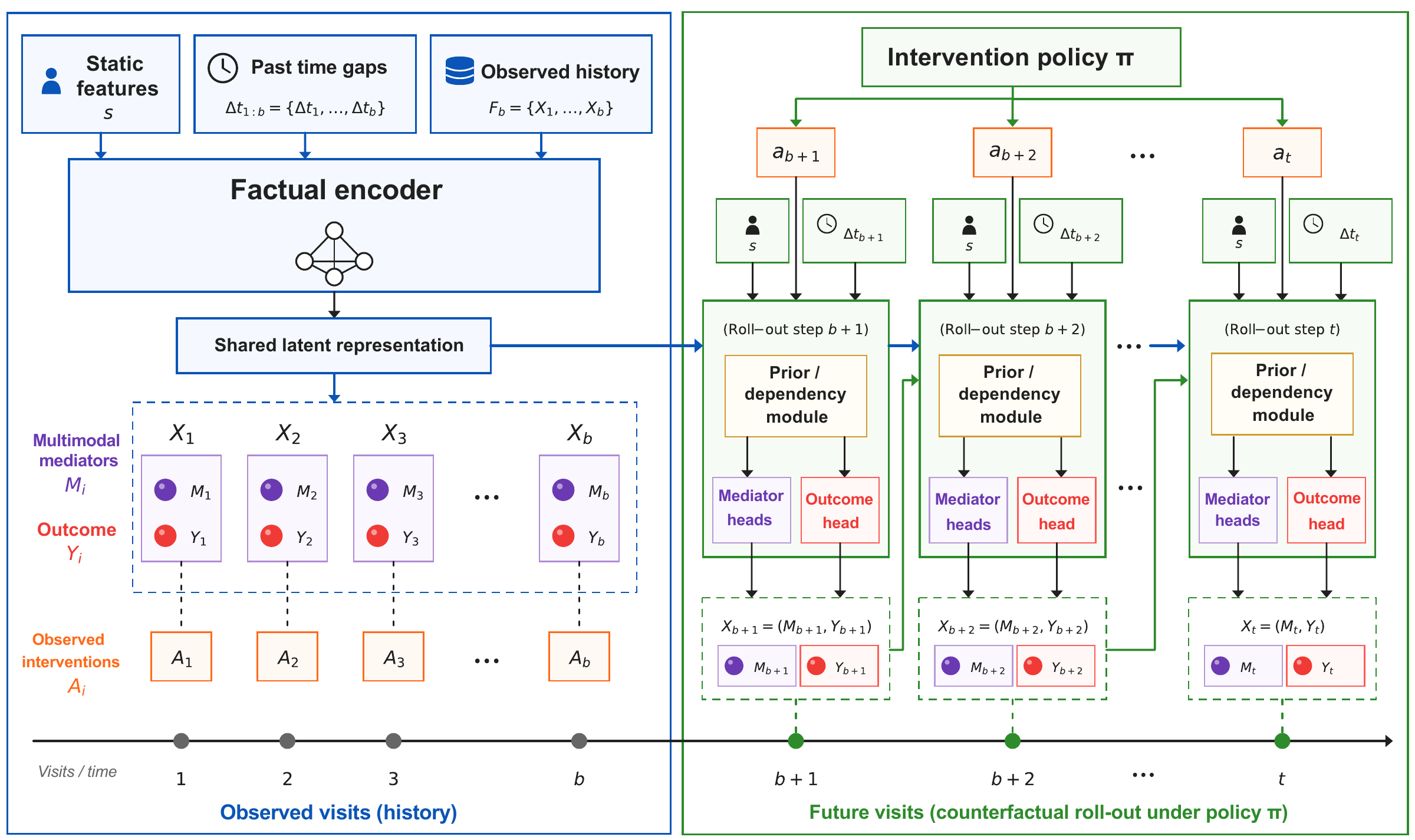}
    \caption{CHDT encodes observed patient history and autoregressively generates future mediator–outcome states.}
    \label{fig:ma}
\end{figure*}

CHDT consists of an encoder and a decoder, as shown in Figure~\ref{fig:ma}. \textbf{Factual encoder:} encodes the observed patient history, static characteristics, and elapsed time intervals into a latent representation. \textbf{Counterfactual decoder:} sequentially generates ICSs conditioned on the observed history and previously generated ICSs given stochastic noise. The decoder uses prior heads to model dependencies among mediators and produce the final predictive distributions.

In the encoder, the observed history, elapsed time, and static patient characteristics are embedded using either recurrent or transformer-based architectures. The decoder follows Algorithm~\ref{algo:chdt_run}, where $DE(\cdot)$ can be implemented as a recurrent neural network equipped with softmax layers for categorical variables and regression layers for continuous variables. In this study, we adopt a simple LSTM-based implementation.

For continuous variables, the model optimizes a mean squared error or Gaussian negative log-likelihood loss. For categorical and binary variables, the model may optimize a cross-entropy loss. Therefore, the overall training loss can be written as:

\begin{equation}
\mathcal{L}
=
\lambda_{C}\mathcal{L}_{C}
+
\lambda_{D}\mathcal{L}_{D},
\label{eq:training-loss}
\end{equation}

where $\mathcal{L}_{C}$ and $\mathcal{L}_{D}$ denote the loss for continuous and categorical variables, respectively.

\begin{algorithm}[t]
\caption{Counterfactual Decoder}
\label{algo:chdt_run}
\begin{algorithmic}[1]
\Require Encoded history $E_b$, decoder $DE(\cdot)$, policy $\pi$, elapsed intervals $\Delta t$, horizon $t$, simulations $J$
\Ensure Candidate counterfactual trajectory set $\mathcal{S}$
\State $\mathcal{S} \leftarrow \emptyset$
\For{$j=1,\ldots,J$}
    \State $\hat{X}_{b+1:t}^{\pi,(j)} \leftarrow \emptyset$
    \For{$k=b+1,\ldots,t$}
        \State $\hat{X}_{k}^{\pi,(j)} \leftarrow DE\!\left(E_b,\hat{X}_{<k}^{\pi,(j)},\Delta t_{<k},a_k^\pi\right)$
        \State $\hat{X}_{b+1:t}^{\pi,(j)} \leftarrow \hat{X}_{b+1:t}^{\pi,(j)} \cup \{\hat{X}_{k}^{\pi,(j)}\}$
    \EndFor
    \State $\mathcal{S} \leftarrow \mathcal{S} \cup \{\hat{X}_{b+1:t}^{\pi,(j)}\}$
\EndFor
\State \Return $\mathcal{S}$
\end{algorithmic}
\end{algorithm}

\subsection{Multi-Step Rollout Error Bound}
\label{sec:rollout_bound}

Because CHDT generates trajectories autoregressively, local errors may
propagate across future steps. Following standard multi-step analyses for
Lipschitz transition models~\citep{asadi2018lipschitz}, we characterize
this effect using an augmented rollout state $Z_k$ containing the
information propagated by the decoder. For fixed $F_b$, $s$, $\pi$,
$\Delta t$, and rollout randomness, let $\mathcal{T}_k$ and
$\hat{\mathcal{T}}_k$ denote the target and learned one-step transition
maps at future step $k$, respectively.

\paragraph{Proposition (multi-step rollout error).}
Assume that, over the evaluated rollout region,
\begin{equation}
\|\hat{\mathcal{T}}_k(z)-\mathcal{T}_k(z)\|_2
\leq \epsilon_k,
\end{equation}
and that $\mathcal{T}_k$ is $L_k$-Lipschitz,
\begin{equation}
\|\mathcal{T}_k(z)-\mathcal{T}_k(z')\|_2
\leq L_k\|z-z'\|_2.
\end{equation}
If the target and learned rollouts start from the same factual state, then
\begin{equation}
\label{eq:rollout_bound}
\|\hat{Z}_t-Z_t\|_2
\leq
\sum_{k=b+1}^{t}
\epsilon_k
\prod_{\ell=k+1}^{t}L_\ell .
\end{equation}
The proof is provided in Appendix~\ref{app:rollout_bound_proof}. This result follows from the triangle inequality under the stated
Lipschitz assumptions. Because we do not estimate the Lipschitz constants
$L_\ell$ or one-step errors $\epsilon_k$ for CHDT, the bound should be
interpreted as a characterization of error propagation rather than a
guarantee of bounded long-horizon rollout error in practice. Empirical
rollout error is reported in Table~\ref{tab:both_rmse}.

\section{Experiments}
For evaluation, we use dementia as the case study because of its long-term, irregular, and multimodal progression. Statin exposure is selected as a clinically relevant but debated intervention benchmark~\citep{olmastroni2022statin}.
\subsection{Experimental Setup}
\label{sec:data}
The experimental study evaluates CHDT under four scenarios in discriminative and generative modes: RIS (Real In-Source evaluation), where models are trained and evaluated within real-world ADNI data; SIS (Synthetic In-Source evaluation), where the same protocol is applied within synthetic ALPACA data; R2S, where models trained on ADNI are evaluated on ALPACA arms; and S2R, where models trained on ALPACA are evaluated on ADNI arms.

\paragraph{Data \& Causal Assumptions}
CHDT is evaluated using two complementary longitudinal dementia datasets: the real-world ADNI cohort\footnote{Data used in this study were obtained from the Alzheimer's Disease Neuroimaging Initiative (ADNI) database (\url{https://adni.loni.usc.edu/}) and downloaded on 5 March 2026. ADNI was launched in 2003 as a public--private partnership led by Michael W. Weiner, MD, to investigate the use of imaging, biomarkers, and clinical assessments in tracking progression of mild cognitive impairment and Alzheimer's disease.}~\citep{petersen2010adni} (Appendix~\ref{Andi}) and a simulator-generated cohort based on ALPACA\footnote{\url{https://github.com/nolanrbrady/ALPACA-RL}}~\citep{brady2026alpaca}. The simulated cohort is generated using ALPACA as the primary medical measurement engine, augmented with a stochastic diagnosis-progression layer to improve clinical realism and reduce label leakage. In this layer, dementia conversion is modeled as a longitudinal event sampled from a hidden progression process, constrained to clinically plausible deterioration patterns (Appendix~\ref{alpaca}).

Each cohort comprises Development, Placebo, and Statin arms. The Development arm is divided into training, validation, and factual test subsets (80/10/10), while Placebo and Statin are reserved for counterfactual evaluation. In ALPACA, each arm contains 2,000 patients, approximately balanced between Normal (N) and Dementia (D). ADNI is more heterogeneous, comprising 4,103 Development patients (2,391 N, 1,292 M, 420 D), 443 Placebo patients (170 N, 169 M, 104 D), and 1,654 Statin patients (1,041 N, 439 M, 174 D), where M denotes mild cognitive impairment (MCI).  Furthermore, statin exposure is encoded at each visit from medication records available for that visit; future
medication use is not used to label earlier treatment states.

\textbf{Consistency.} We assume consistency between the observed outcomes and the corresponding recorded treatment histories. Treatment exposure is defined at each visit using the medication information available at that visit.
\textbf{Positivity.} We assess empirical treatment support through the presence of both statin-exposed and non-statin trajectories and further examine overlap using propensity-score diagnostics. \textbf{Sequential exchangeability.} We assume sequential exchangeability conditional on the observed longitudinal history. Rich multimodal covariates are included to make this assumption more plausible; however, residual unmeasured confounding cannot be excluded in ADNI. In ALPACA, treatment and disease progression are governed by the specified data-generation mechanism, providing a controlled setting in which intervention-conditioned trajectories can additionally be evaluated. 
\paragraph{Baselines} We compare CHDT with CRN, GTrans, TECDE, RMSN, and GNET introduced in the Related Work section. Since almost all baselines are outcome-prediction models, with the exception of GNET, comparisons are restricted to outcome-level fidelity, while CHDT and GNET are additionally evaluated for full mediator--outcome trajectory generation.
\paragraph{Metrics}
Trajectory fidelity is evaluated on active patient--visit entries using classification metrics, together with propensity-score-based treatment-support diagnostics.
\paragraph{Implementation Details}
For all models, the weight decay is set to 0, the number of training epochs to 50, the batch size is fixed at 64, and the Adam optimizer is used. Key hyperparameters, including the dropout rate ($\{0.1, 0.2\}$), learning rate ($\{0.01, 0.001, 0.0001\}$), and number of layers ($\{1, 2, 3\}$), with the number of Monte Carlo samples ($\{1, 3, 5, 10\}$), are selected through an exhaustive grid search over the predefined search space\footnote{Each baseline is tuned independently using the same validation split and comparable search budget, 
with method-specific hyperparameters selected according to the original implementation where applicable. Additional pre-processing, architecture, and hyperparameter details are provided in the accompanying code repository.}. Each experiment is run 10 times on a Red Hat Enterprise Linux system with 768 GB of memory, using Python 3.12.6 and PyTorch 2.11.0 with CUDA 13.0.

\subsection{Sensitivity Analysis}
In the RIS setting, sensitivity analysis of the loss weights ($[\lambda_D,\lambda_C]$) across several configurations with values ranging from 1 to 3 shows only modest performance variation. Increasing the continuous-loss weight generally improves performance, with ($[1,3]$) achieving the highest mean Macro F1 scores of 0.8530 (mean) with var ($4.15\times10^{-4}$) on the validation set. We also evaluated sensitivity to the Monte Carlo sample size ($J \in \{1,3,5,10\}$). Performance varied only slightly with $J$, with Macro F1 changing by approximately 0.01 on the validation set.

\subsection{Ablation Study}

To assess the contribution of each component, we conducted an ablation study in the RIS setting. Baseline+T incorporates irregular time intervals, whereas Baseline+C introduces cross-mediator dependency modeling; CHDT combines both components. As reported in Table~\ref{tab:gnet_ablation_realworld_adni}, temporal modeling substantially improves over the Baseline on Test (Micro/Macro F1: $+0.064/+0.083$) and Placebo ($+0.071/+0.072$). Cross-mediator modeling also improves performance across all three splits. Overall, the ablation results indicate that explicit modeling of irregular visit intervals provides the largest performance gain, while cross-mediator refinement provides additional improvements when combined with temporal modeling.

\begin{table*}[h]
\centering
\begingroup
\fontsize{9pt}{11pt}\selectfont
\resizebox{\textwidth}{!}{%
\begin{tabular}{@{}lcccccc@{}}
\toprule
 & \multicolumn{2}{c}{Test} & \multicolumn{2}{c}{Placebo} & \multicolumn{2}{c}{Statin} \\
\cmidrule(lr){2-3}\cmidrule(lr){4-5}\cmidrule(l){6-7}
Model & Micro F1 & Macro F1 & Micro F1 & Macro F1 & Micro F1 & Macro F1 \\
\midrule
Baseline
 & 0.719 ($9.47\times10^{-5}$) & 0.636 ($2.60\times10^{-4}$)
 & 0.584 ($1.28\times10^{-4}$) & 0.541 ($1.38\times10^{-4}$)
 & 0.489 ($1.67\times10^{-4}$) & 0.375 ($7.93\times10^{-4}$) \\
Baseline+T
 & 0.783 ($1.66\times10^{-4}$) & 0.719 ($1.92\times10^{-4}$)
 & 0.655 ($1.11\times10^{-4}$) & 0.613 ($1.30\times10^{-4}$)
 & 0.471 ($4.41\times10^{-4}$) & \textbf{0.385 ($7.95\times10^{-4}$)} \\
Baseline+C
 & 0.759 ($3.57\times10^{-5}$) & 0.682 ($3.81\times10^{-5}$)
 & 0.617 ($1.65\times10^{-4}$) & 0.573 ($1.53\times10^{-4}$)
 & 0.494 ($2.85\times10^{-4}$) & 0.381 ($6.62\times10^{-4}$) \\
CHDT
 & \textbf{0.830 ($2.61\times10^{-5}$)} & \textbf{0.768 ($4.66\times10^{-5}$)}
 & \textbf{0.679 ($8.48\times10^{-5}$)} & \textbf{0.663 ($9.38\times10^{-5}$)}
 & \textbf{0.497 ($3.07\times10^{-4}$)} & 0.384 ($6.43\times10^{-4}$) \\
\bottomrule
\end{tabular}%
}
\caption{Ablation results for the Baseline, Baseline+T, Baseline+C, and CHDT averaged over
horizons 1--4 (mean--variance across 10 runs). }
\label{tab:gnet_ablation_realworld_adni}
\endgroup
\end{table*}

\subsection{Factual Checks}

Factual checks assess whether CHDT can generate faithful observed trajectories, which is essential because counterfactual estimation is meaningful only after the model captures factual longitudinal disease evolution.

\subsubsection{Factual Fidelity Comparison}

\begin{figure*}[t]
    \centering
    \includegraphics[width=\linewidth]{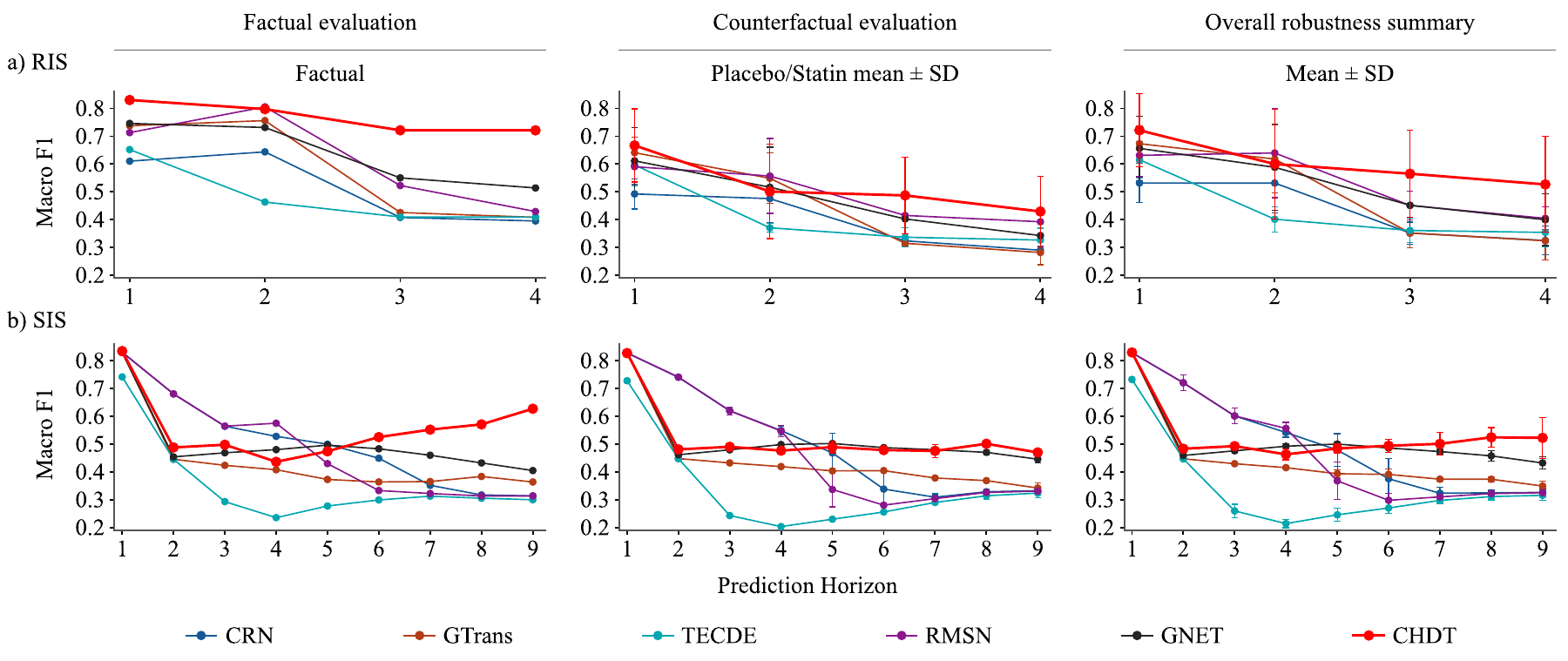}
    \caption{Comparison of factual and counterfactual predictive performance across horizons in RIS and SIS.}
    \label{fig:gnetpp_sis_macro_f1}
\end{figure*}

RIS is the most clinically relevant factual setting, as the data include irregular time intervals and class imbalance. Figure~\ref{fig:gnetpp_sis_macro_f1} indicates that CHDT generally achieves more consistent classification performance than the compared baselines. The horizon-wise results in Appendix~\ref{app:RIS_horizon_results}, Tables~\ref{tab:macro_f1_adni} and~\ref{tab:dementia_f1_adni},  show that CHDT achieves the highest mean Macro F1 at three of the four factual horizons and the highest mean Dementia F1 at all four horizons. At the 2-step horizon, RMSN achieves a marginally higher Macro F1 than CHDT (0.8057 vs. 0.7977).

CHDT achieves mean Dementia F1 scores of 0.738, 0.752, 0.535, and 0.646 from one-step to four-step prediction. Its normalized Macro F1 drop is 13.15\%, which is lower than those of the compared baselines, as reported in Table~\ref{tab:horizon_degradation}. In SIS, the factual panel in Figure~\ref{fig:gnetpp_sis_macro_f1} shows that CHDT achieves favorable Macro F1 performance across prediction horizons. Although some baselines are competitive at early horizons, their performance declines more over longer horizons. CHDT has the smallest normalized factual Macro F1 drop in this setting, at 24.84\%.

Overall, in irregular real-world ADNI trajectories, CHDT maintains strong disease-state performance over
autoregressive horizons, providing the primary evidence that the proposed temporal formulation improves longitudinal
fidelity. Additional horizon-wise results are reported in Appendix~\ref{app:RIS_horizon_results}, Tables~\ref{tab:macro_f1_adni}--\ref{tab:dementia_f1_pvalues_adni}.

\begin{table}[!ht]
\centering
\begingroup
\fontsize{8pt}{8pt}\selectfont
\setlength{\tabcolsep}{3pt}
\begin{tabular}{llcc}
\toprule
Model & Arm & RIS & SIS\\
\midrule
CRN & Test & $35.25\;(1.5\!\times\!10^{1})$ & $62.23\;(2.0\!\times\!10^{-4})$ \\
 & Placebo & $44.37\;(1.3\!\times\!10^{1})$ & $61.57\;(7.9\!\times\!10^{-1})$ \\
 & Statin & $37.04\;(3.4\!\times\!10^{2})$ & $58.13\;(3.7\!\times\!10^{-2})$ \\
\addlinespace
GTrans & Test & $44.66\;(1.0\!\times\!10^{1})$ & $56.23\;(2.1\!\times\!10^{1})$ \\
 & Placebo & $53.27\;(3.5\!\times\!10^{0})$ & $56.61\;(3.1\!\times\!10^{1})$ \\
 & Statin & $59.34\;(1.8\!\times\!10^{2})$ & $60.87\;(8.0\!\times\!10^{0})$ \\
\addlinespace
TECDE & Test & $37.28\;(6.2\!\times\!10^{1})$ & $59.56\;(4.6\!\times\!10^{-2})$ \\
 & Placebo & $57.64\;(1.0\!\times\!10^{1})$ & $58.29\;(6.2\!\times\!10^{-2})$ \\
 & Statin & $29.40\;(4.5\!\times\!10^{2})$ & $52.69\;(6.5\!\times\!10^{-2})$ \\
\addlinespace
RMSN & Test & $39.81\;(3.5\!\times\!10^{0})$ & $62.24\;(5.6\!\times\!10^{-29})$ \\
 & Placebo & $46.89\;(1.0\!\times\!10^{1})$ & $61.86\;(1.8\!\times\!10^{-1})$ \\
 & Statin & $\mathbf{17.01}\;(4.8\!\times\!10^{2})$ & $58.19\;(1.2\!\times\!10^{-4})$ \\
\addlinespace
GNET & Test & $31.10\;(2.5\!\times\!10^{1})$ & $51.22\;(1.0\!\times\!10^{2})$ \\
 & Placebo & $45.39\;(2.4\!\times\!10^{1})$ & $48.09\;(9.9\!\times\!10^{0})$ \\
 & Statin & $41.84\;(1.5\!\times\!10^{2})$ & $44.20\;(3.7\!\times\!10^{1})$ \\
\addlinespace
CHDT & Test & $\mathbf{13.15}\;(1.7\!\times\!10^{1})$ & $\mathbf{24.84}\;(1.1\!\times\!10^{-1})$ \\
 & Placebo & $\mathbf{30.33}\;(5.5\!\times\!10^{0})$ & $\mathbf{43.15}\;(1.7\!\times\!10^{0})$ \\
 & Statin & $43.52\;(1.7\!\times\!10^{1})$ & $\mathbf{43.19}\;(1.8\!\times\!10^{1})$ \\
\bottomrule
\end{tabular}
\endgroup
\caption{Classification stability across all arms.
RIS and SIS percentage drops are calculated from the reported
mean and variance of the Macro F1 scores.}
\label{tab:horizon_degradation}
\end{table}

\subsubsection{Factual Cross-Source Fidelity}
\begin{figure}[t]
\centering
\includegraphics[width=0.9\linewidth]{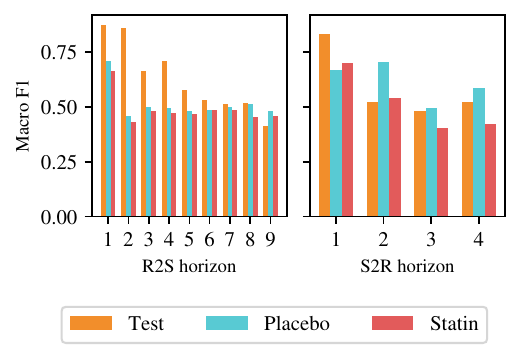}
\caption{Horizon stability of CHDT in R2S and S2R.}
\label{fig:r2s_s2r_autoregressive_horizon_stability}
\end{figure}

In R2S, as shown in Figure~\ref{fig:r2s_s2r_autoregressive_horizon_stability}, CHDT retains factual prediction capability when trained on ADNI and evaluated on ALPACA, suggesting that the learned dynamics capture some transferable structure of disease progression. Similarly, the S2R results suggest that synthetic training captures patterns that support factual prediction in ADNI despite its irregularity, sparsity, and observational complexity.

\subsection{Counterfactual Checks}

We assess intervention-conditioned fidelity by comparing generated trajectories with held-out observed treatment arms. These evaluations provide indirect counterfactual checks.
\subsubsection{Intervention-Conditioned Fidelity Comparison}

In RIS, the counterfactual panel of Figure~\ref{fig:gnetpp_sis_macro_f1} shows favorable overall mean performance for CHDT, particularly at later horizons. The arm-specific results show the clearest advantage in the Placebo arm. CHDT obtains the highest mean Macro F1 at three of the four Placebo horizons and the highest mean Dementia F1 at all four horizons, as reported in Appendix~\ref{app:RIS_horizon_results}, Tables~\ref{tab:macro_f1_adni} and~\ref{tab:dementia_f1_adni}. Its Macro F1 decreases from 0.799 at step 1 to 0.557 at step 4, with a normalized drop of 30.33\%, the smallest among the compared models in this arm (Table~\ref{tab:horizon_degradation}). Averaging equally across both RIS intervention arms and all horizons, CHDT also obtains the highest mean Macro F1, at 0.521. These results suggest favorable overall disease-state preservation, particularly under the Placebo policy.

In SIS, the counterfactual panel of Figure~\ref{fig:gnetpp_sis_macro_f1} shows that CHDT maintains favorable mean performance across the two intervention arms at later horizons, although some baselines perform better at earlier horizons. In particular, the arm-specific results in Table~\ref{tab:horizon_degradation} show that CHDT achieves the smallest normalized Macro F1 drop in the Statin arm, at 43.19\%, compared with 44.20--60.87\% for the baselines. This suggests better retention of classification performance under the Statin policy over longer rollouts, although the difference from GNET is modest. CHDT also obtains the smallest normalized drop in the Placebo arm, at 43.15\%, indicating favorable horizon stability across both synthetic intervention arms.

The exploratory two-sided Welch $t$-tests against CHDT are reported in Appendix~\ref{app:RIS_horizon_results}, Tables~\ref{tab:macro_f1_pvalues_adni} and~\ref{tab:dementia_f1_pvalues_adni}. Most RIS Placebo comparisons yield ($p<0.01$) in favor of CHDT for Macro F1 and Dementia F1, providing complementary evidence of improved performance across runs.

Overall, CHDT shows favorable mean counterfactual classification performance in RIS, particularly in the Placebo arm. The overall robustness summary in Figure~\ref{fig:gnetpp_sis_macro_f1} is consistent with this favorable later-horizon performance across factual and counterfactual evaluations. Agreement with observed ADNI treatment-arm outcomes provides an indirect fidelity check rather than direct validation of individual counterfactual trajectories.

\subsubsection{Cross-Source Counterfactual Robustness}

The cross-source counterfactual check evaluates whether predictions of counterfactual trajectories remain stable when training and evaluation are performed across different sources. This setting provides an additional counterfactual robustness check.

As shown in Figure~\ref{fig:r2s_s2r_autoregressive_horizon_stability}, in R2S, CHDT maintains Macro F1 values around 0.50 across most horizons, with the first horizon exceeding 0.65 in the Placebo and Statin
counterfactual results. This trend is broadly consistent with the SIS counterfactual setting in Figure~\ref{fig:gnetpp_sis_macro_f1}, where CHDT also performs around 0.50. This suggests that CHDT preserves a comparable level of counterfactual prediction behavior even under cross-source evaluation.

Overall, R2S serves as a useful robustness test for assessing whether patterns learned from ADNI transfer to the controlled ALPACA simulation setting.

\begin{figure}[t]
\centering
\includegraphics[width=0.8\linewidth]{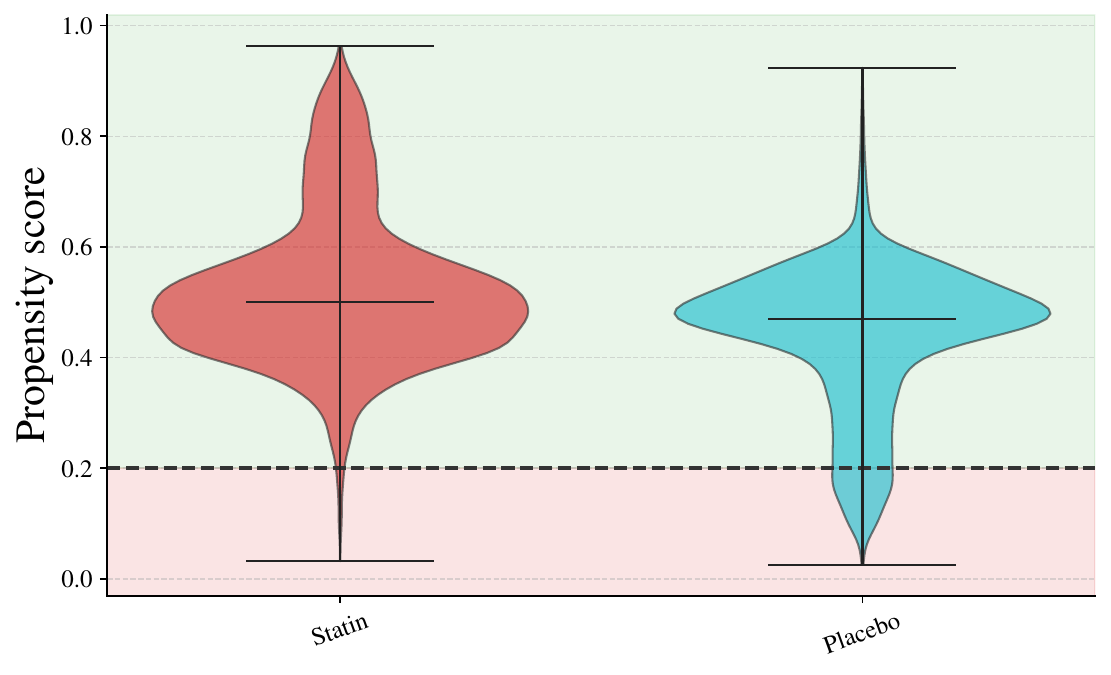}
\caption{Propensity scores across generated Placebo and Statin trajectories in the R2S setting: $e(H_t)$ for Statin and $1-e(H_t)$ for Placebo.}
\label{fig:assigned_treatment_support_by_dataset}
\end{figure}

\subsubsection{Counterfactual Plausibility}
The propensity score $e(H_t)$ estimates the probability of Statin exposure for a given patient history under the real-world ADNI distribution. This assesses whether generated counterfactual trajectories are supported by real-world treatment patterns, while the overlap violation rate measures the proportion of trajectories with propensity scores below the threshold $\alpha$~\citep{rosenbaum1983central,hernan2020causal}.
To estimate the propensity scores, a Maximum Entropy (MaxEnt) model was trained and optimized on ADNI and evaluated on counterfactual trajectories generated in R2S. Before applying the treatment-support diagnostic to generated trajectories, we verified the propensity
model on held-out ADNI data and investigated its discrimination and calibration. In addition, the model is conditioned only on patient history available prior to the action at visit
$t$.

Figure~\ref{fig:assigned_treatment_support_by_dataset} demonstrates the propensity score for each dataset and intervention arm for the R2S setting, where the dashed horizontal line indicates the overlap threshold $\alpha=0.2$, below which trajectories are considered weakly supported.

The results show that most generated patient-state trajectories remain above the overlap threshold, suggesting that R2S counterfactual trajectories do not appear to be dominated by unsupported extrapolation from the real-world ADNI distribution. Both generated Statin and Placebo patient-state trajectories mainly fall within intermediate propensity scores, around 0.40--0.55, indicating that they lie in a central overlap region where both treatment assignments appear plausible.
\subsection{Trajectory Generation Evaluation}
In RIS, CHDT and GNET are evaluated in generative mode using normalized RMSE, obtained by normalizing each feature's prediction error by its observed range, computing RMSE separately for each feature, and averaging across features. As shown in Table~\ref{tab:both_rmse}, CHDT achieves a lower Total RMSE than GNET (0.185 vs.\ 0.194; $p<0.01$) and consistently lower errors across horizons 1--4 (0.179--0.210 vs.\ 0.183--0.216), indicating more accurate and stable autoregressive trajectory generation. As further detailed in Appendix~\ref{app:feature_rmse}, feature-wise analysis shows that CHDT generally achieves lower errors than GNET across cognitive, biomarker, and MRI-derived variables, while CDR-SB remains the most challenging feature for both models. 

\begin{table}[t]
\centering
\caption{Horizon-wise (both arms) normalized RMSE comparison between CHDT and GNET (mean--variance across 10 runs; two-sided $t$-test: $p<0.01$ for all horizons and Total).}
\label{tab:both_rmse}
\begingroup
\fontsize{8pt}{9pt}\selectfont
\setlength{\tabcolsep}{5pt}
\resizebox{\columnwidth}{!}{%
\begin{tabular}{ccc}
\toprule
Horizon & CHDT RMSE & GNET RMSE \\
\midrule
1     & $\mathbf{0.179}\;(5.08\times10^{-6})$ & $0.183\;(4.43\times10^{-6})$ \\
2     & $\mathbf{0.174}\;(1.38\times10^{-5})$ & $0.184\;(1.04\times10^{-5})$ \\
3     & $\mathbf{0.185}\;(3.21\times10^{-5})$ & $0.200\;(1.56\times10^{-5})$ \\
4     & $\mathbf{0.210}\;(1.62\times10^{-5})$ & $0.216\;(1.84\times10^{-5})$ \\
\midrule
Total & $\mathbf{0.185}\;(5.69\times10^{-6})$
      & $0.194\;(1.87\times10^{-6})$ \\
\bottomrule
\end{tabular}%
}
\endgroup
\end{table}

\section{Conclusion}
This paper introduced CHDT, an intervention-conditioned Human Digital Twin generator for multimodal longitudinal trajectory estimation. Across real-world and simulator-generated cohorts, the results indicate that explicit irregular-time conditioning provides the primary improvement in long-horizon patient-state fidelity, with additional gains from cross-mediator refinement. CHDT also produces intervention-conditioned trajectories that largely remain within empirically supported treatment regions; however, these analyses provide indirect counterfactual evidence rather than direct validation of individual potential outcomes.

\section*{Acknowledgements}
This research was supported by the National Institute for Health and Care Research (NIHR).  RA is supported by the NIHR Southampton Biomedical Research Centre. SM is supported by an NIHR Senior Clinical Practitioner Award. 

Data collection and sharing for the Alzheimer's Disease Neuroimaging Initiative (ADNI) is funded by the National Institute on Aging (National Institutes of Health Grant U19AG024904) and through generous contributions from private sector and non-profit sponsors\footnotetext{Data used in preparation of this article were obtained from the Alzheimer's Disease Neuroimaging Initiative (ADNI) database (\url{https://adni.loni.usc.edu/}). As such, the investigators within ADNI contributed to the design and implementation of ADNI and/or provided data but did not participate in the analysis or writing of this report. A complete listing of ADNI investigators can be found at \url{https://adni.loni.usc.edu/wp-content/uploads/how_to_apply/ADNI_Acknowledgement_List.pdf}}.

\bibliography{myreferences}

\clearpage

\appendix

\section{Proof of the Multi-Step Rollout Error Bound}
\label{app:rollout_bound_proof}

Let $e_k=\|\hat{Z}_k-Z_k\|_2$. Using
$\hat{Z}_k=\hat{\mathcal{T}}_k(\hat{Z}_{k-1})$ and
$Z_k=\mathcal{T}_k(Z_{k-1})$, the triangle inequality gives
\begin{align}
e_k
&\leq
\|\hat{\mathcal{T}}_k(\hat{Z}_{k-1})
-\mathcal{T}_k(\hat{Z}_{k-1})\|_2 \nonumber\\
&\quad+
\|\mathcal{T}_k(\hat{Z}_{k-1})
-\mathcal{T}_k(Z_{k-1})\|_2 \nonumber\\
&\leq \epsilon_k+L_k e_{k-1}.
\end{align}
Since the rollouts share the same factual state, $e_b=0$.
Recursively applying the inequality yields
\[
e_t
\leq
\sum_{k=b+1}^{t}
\epsilon_k
\prod_{\ell=k+1}^{t}L_\ell,
\]
where an empty product equals $1$, proving
Eq.~\eqref{eq:rollout_bound}. \hfill$\square$

\section{Data}
\subsection{Real-world ADNI Cohort}
\label{Andi}
The ADNI cohort was constructed as a visit-level longitudinal dataset indexed by participant and visit code. Clinical, demographic, cognitive, CSF biomarker, MRI-volume, and medication records were harmonized into a single longitudinal table. Medication records were aggregated to the visit level before integration.
\subsection{Synthetic Longitudinal Data Generation: ALPACA Cohort}
\label{alpaca}
The synthetic cohort was generated using the Alzheimer's Learning Platform for Adaptive Care Agents (ALPACA)~\citep{brady2026alpaca}. The real-world cohort was extracted from ADNI using the Analysis Ready Cohort (ARC) Builder and subsequently integrated with MRI-volume and medication information.

In this study, the simulator was configured using the impaired cohort setting and the \texttt{ADNI\_MEM} reward metric. Observed cognitive and MRI deterioration restricts the set of clinically plausible conversion visits,
while the final conversion event and timing are sampled stochastically from a hidden progression process. Treatment assignment was defined through three generated arms: placebo, sustained statin treatment, and a mixed arm in which statin exposure could begin during follow-up.

Synthetic progressive longitudinal dementia trajectories were generated using ALPACA as the primary medical measurement engine, together with an additional stochastic diagnosis-progression layer designed to improve clinical realism and reduce label leakage. ALPACA was first used to simulate patient-level longitudinal trajectories containing demographic variables, cognitive assessments, CSF biomarkers, MRI-derived brain volumes, treatment exposure, and treatment-conditioned temporal progression patterns. The generated dataset was stored in long format, where each row represents one patient visit and repeated visits form a longitudinal disease trajectory suitable for autoregressive HDT training. For the placebo arm, patients received no active medication during follow-up, while their clinical measurements evolved according to the ALPACA transition dynamics.

To avoid unrealistic patient-level fixed dementia labels, a new stochastic diagnosis layer was introduced on top of the ALPACA-generated measurements. In the proposed generator, dementia conversion is treated as an incident longitudinal clinical event rather than a predefined patient identity. All patients begin with normal status at baseline, and only a subset of patients are selected as potential converters during follow-up. For these patients, the conversion visit is not determined using a deterministic threshold on observed biomarkers. Instead, conversion timing is sampled probabilistically from a hidden stochastic disease progression process containing latent patient frailty, progression rate, random innovation noise, age-related risk, treatment exposure, and prior clinical worsening. Consequently, patients with similar observed biomarker values may convert at different visits, while some patients may not convert during the observed follow-up window, reflecting the uncertainty and heterogeneity commonly observed in real clinical dementia progression.

Importantly, the diagnosis layer operates under clinical plausibility constraints. A visit becomes eligible for dementia conversion only when the surrounding ALPACA measurements indicate plausible deterioration patterns, such as decline in ADNI\_MEM or ADNI\_EF2 scores, ventricular enlargement, or hippocampal and entorhinal volume reduction. However, these measurements are not used as direct deterministic rules for assigning the dementia label. Instead, they only constrain the set of clinically plausible conversion visits, while the final conversion decision remains stochastic and dependent on the hidden latent severity trajectory. Once conversion occurs, all subsequent visits remain labeled $AD=1$, producing monotonic progressive dementia trajectories.

This additional stochastic progression layer plays an important role in improving the realism and validity of the generated dataset. Hidden variables such as frailty, latent severity progression, and conversion targets are never exported to the final dataset, thereby reducing information leakage. Future visit information is not used for current-visit prediction, and dementia status is no longer constant across all visits for a patient. Instead, dementia trajectories contain observable transitions from $AD=0$ to $AD=1$ with progressive but noisy clinical deterioration. This design produces more realistic longitudinal disease trajectories with uncertainty, overlap, and heterogeneous progression patterns that better resemble real-world dementia evolution.

\subsection{Data Harmonization}
\label{sec:data_harmonization}
To enable cross-source transfer evaluation between the simulator-generated ALPACA cohort and the real-world ADNI cohort under the R2S and S2R scenarios, a feature-level harmonization procedure was performed to construct a shared longitudinal representation across both datasets. Harmonization was defined as retaining only clinically comparable variables that were available in both sources while removing source-specific features. The final harmonized representation included patient identifier, visit index, visit gap, age, gender, treatment exposure, CSF biomarkers (ABETA and TAU), MRI-derived anatomical measurements (hippocampus, entorhinal cortex, and fusiform volumes), and the dementia outcome variable. To ensure schema consistency during cross-source training and evaluation, all harmonized outputs were converted into a unified ADNI-style representation using common column names. Variables without reliable cross-source correspondence were excluded from the harmonized space. In particular, ventricle-related variables were removed because the ALPACA dataset contains a single aggregated ventricle measurement whereas ADNI provides separate third- and fourth-ventricle volumes with substantially different empirical distributions.

\subsection{Feature-wise Generative Fidelity}
\label{app:feature_rmse}

We further assess generative fidelity by decomposing the Total normalized RMSE across individual patient-state features. Prediction errors are normalized by each feature's observed range, RMSE is computed separately for each feature, and results are averaged across runs. The evaluated variables include CDR-SB (global cognitive and functional impairment), MMSE and TOTAL13 (cognitive performance), ABETA42 and TAU (biomarkers), and MRI-derived hippocampal, entorhinal, fusiform, third-ventricle, and fourth-ventricle volumes.

\begin{figure}[t]
    \centering
    \includegraphics[width=\columnwidth]{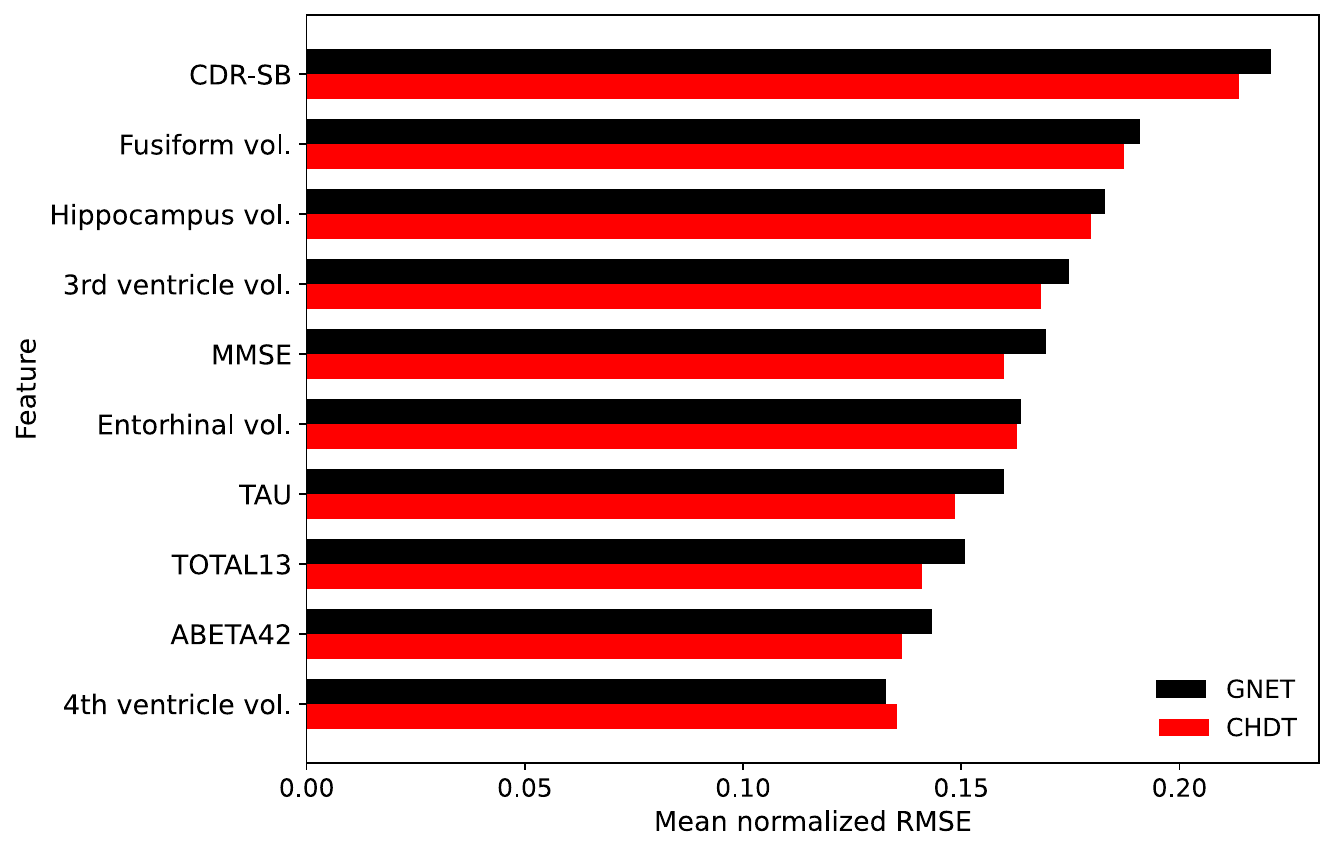}
    \caption{Feature-wise normalized RMSE for CHDT and GNET using the Total generative evaluation. Lower values indicate better generative fidelity.}
    \label{fig:feature_rmse}
\end{figure}

\begin{figure}[t]
    \centering
    \includegraphics[width=0.8\linewidth]{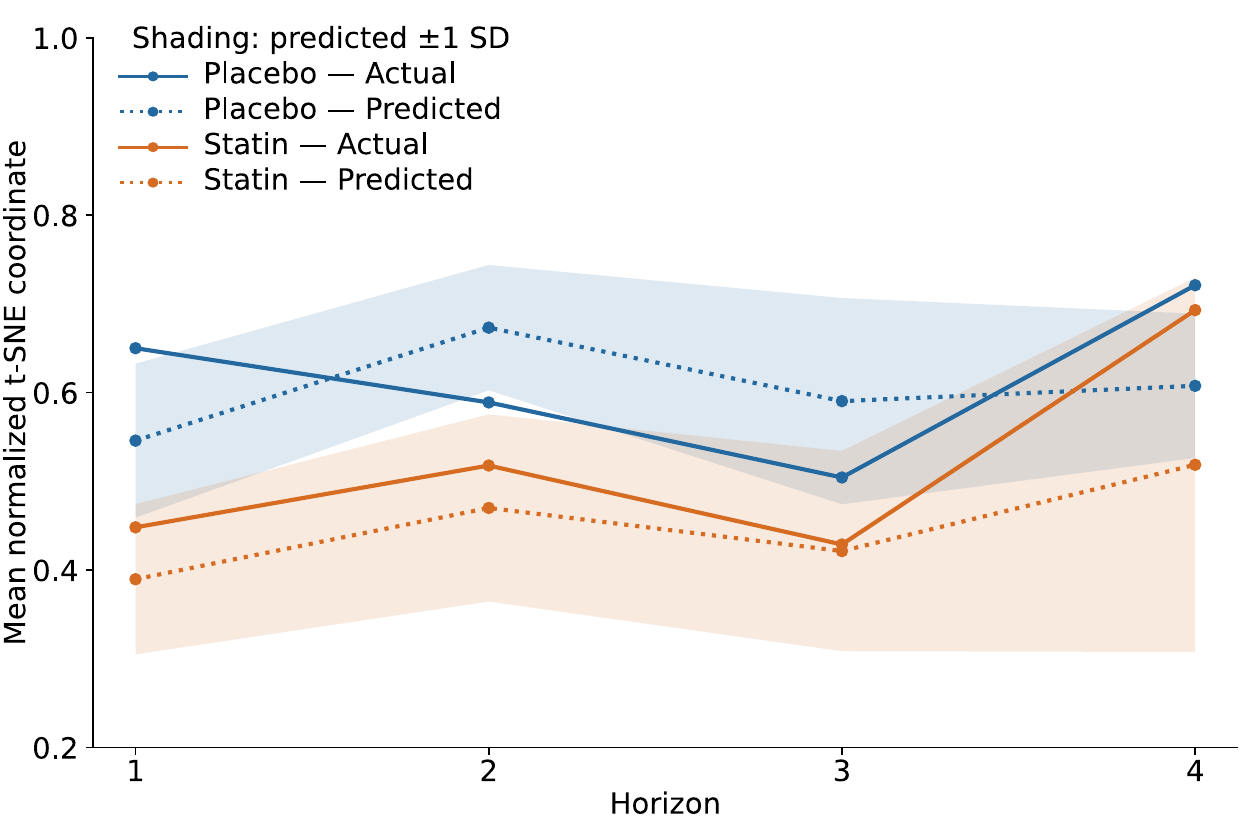}
    \caption{Actual and CHDT-generated patient-state trajectories in RIS.}
    \label{fig:patient_state_trajectories}
\end{figure}

As shown in Figure~\ref{fig:feature_rmse}, CHDT achieves lower RMSE than GNET for 9 of the 10 evaluated features, suggesting that its overall improvement is distributed across heterogeneous components of the generated patient state rather than being driven by a single variable. The largest reductions are observed for TAU (0.148 vs.\ 0.160), TOTAL13 (0.141 vs.\ 0.151), and MMSE (0.160 vs.\ 0.169). CDR-SB remains the most challenging feature for both models (0.214 vs.\ 0.221), while performance is nearly identical for entorhinal volume. GNET is marginally better only for fourth-ventricle volume (0.133 vs.\ 0.135). Overall, CHDT shows consistently improved generative fidelity across cognitive, biomarker, and structural MRI variables. These results evaluate generated trajectory fidelity at the feature level rather than calibration of the complete predictive distribution.

Figure~\ref{fig:patient_state_trajectories} shows that CHDT preserves the relative ordering of the two intervention arms and broadly follows the observed Statin trajectory, although some underestimation is visible at horizon 4. Because t-SNE coordinates ~\footnote{This denotes the average position of points in a t-SNE~\citep{van2008visualizing} plot after its coordinates have been normalized.} are embedding-dependent, this analysis is used only as a complementary visualization rather than a quantitative fidelity metric.

\section{Additional RIS Horizon-Wise Results}
\label{app:RIS_horizon_results}

\begin{table*}[p]
\centering
\begingroup
\fontsize{9pt}{9pt}\selectfont
\setlength{\tabcolsep}{3pt}

\begin{tabular}{llcccc}
\toprule
Model & Setting & 1-step & 2-step & 3-step & 4-step \\
\midrule
CRN & Test & $0.6105\;(1.1\!\times\!10^{-3})$ & $0.6439\;(2.5\!\times\!10^{-3})$ & $0.4079\;(3.5\!\times\!10^{-4})$ & $0.3953\;(4.4\!\times\!10^{-5})$ \\
 & Placebo & $0.5466\;(1.3\!\times\!10^{-3})$ & $0.5629\;(2.4\!\times\!10^{-3})$ & $0.3260\;(5.1\!\times\!10^{-4})$ & $0.3041\;(3.4\!\times\!10^{-5})$ \\
 & Statin & $0.4384\;(4.5\!\times\!10^{-3})$ & $0.3884\;(3.9\!\times\!10^{-4})$ & $0.3218\;(1.8\!\times\!10^{-3})$ & $0.2760\;(5.3\!\times\!10^{-3})$ \\
\addlinespace
GTrans & Test & $0.7380\;(1.8\!\times\!10^{-4})$ & $0.7563\;(8.6\!\times\!10^{-3})$ & $0.4257\;(1.5\!\times\!10^{-4})$ & $0.4084\;(2.1\!\times\!10^{-3})$ \\
 & Placebo & $0.6967\;(4.0\!\times\!10^{-4})$ & $0.6398\;(6.5\!\times\!10^{-3})$ & $0.3088\;(9.8\!\times\!10^{-5})$ & $0.3256\;(2.8\!\times\!10^{-3})$ \\
 & Statin & $\mathbf{0.5854}\;(4.1\!\times\!10^{-4})$ & $\mathbf{0.4579}\;(2.7\!\times\!10^{-3})$ & $0.3211\;(4.9\!\times\!10^{-5})$ & $0.2380\;(4.9\!\times\!10^{-4})$ \\
\addlinespace
TECDE & Test & $0.6521\;(3.3\!\times\!10^{-4})$ & $0.4634\;(8.4\!\times\!10^{-5})$ & $0.4097\;(1.9\!\times\!10^{-4})$ & $0.4090\;(3.5\!\times\!10^{-4})$ \\
 & Placebo & $0.6712\;(1.7\!\times\!10^{-4})$ & $0.3544\;(2.8\!\times\!10^{-4})$ & $0.3036\;(1.2\!\times\!10^{-4})$ & $0.2843\;(1.3\!\times\!10^{-4})$ \\
 & Statin & $0.5228\;(1.8\!\times\!10^{-3})$ & $0.3871\;(5.6\!\times\!10^{-4})$ & $0.3700\;(1.5\!\times\!10^{-3})$ & $0.3691\;(4.5\!\times\!10^{-3})$ \\
\addlinespace
RMSN & Test & $0.7132\;(3.5\!\times\!10^{-5})$ & $\mathbf{0.8057}\;(2.1\!\times\!10^{-4})$ & $0.5228\;(3.3\!\times\!10^{-4})$ & $0.4293\;(4.7\!\times\!10^{-5})$ \\
 & Placebo & $0.6532\;(6.2\!\times\!10^{-5})$ & $\mathbf{0.6918}\;(4.8\!\times\!10^{-4})$ & $0.4148\;(5.7\!\times\!10^{-5})$ & $0.3469\;(6.1\!\times\!10^{-5})$ \\
 & Statin & $0.5278\;(3.1\!\times\!10^{-4})$ & $0.4224\;(1.9\!\times\!10^{-4})$ & $\mathbf{0.4156}\;(7.2\!\times\!10^{-5})$ & $\mathbf{0.4380}\;(9.2\!\times\!10^{-4})$ \\
\addlinespace
GNET & Test & $0.7462\;(1.2\!\times\!10^{-4})$ & $0.7316\;(1.2\!\times\!10^{-4})$ & $0.5503\;(1.0\!\times\!10^{-3})$ & $0.5141\;(1.4\!\times\!10^{-3})$ \\
 & Placebo & $0.7307\;(1.3\!\times\!10^{-4})$ & $0.6607\;(3.8\!\times\!10^{-4})$ & $0.4879\;(5.4\!\times\!10^{-4})$ & $0.3990\;(3.0\!\times\!10^{-4})$ \\
 & Statin & $0.4924\;(4.3\!\times\!10^{-4})$ & $0.3727\;(5.8\!\times\!10^{-4})$ & $0.3173\;(3.8\!\times\!10^{-3})$ & $0.2864\;(1.4\!\times\!10^{-3})$ \\
\addlinespace
CHDT & Test & $\mathbf{0.8305}\;(2.0\!\times\!10^{-5})$ & $0.7977\;(2.6\!\times\!10^{-4})$ & $\mathbf{0.7216}\;(3.3\!\times\!10^{-4})$ & $\mathbf{0.7213}\;(1.9\!\times\!10^{-4})$ \\
 & Placebo & $\mathbf{0.7988}\;(5.9\!\times\!10^{-5})$ & $0.6706\;(1.1\!\times\!10^{-4})$ & $\mathbf{0.6243}\;(4.4\!\times\!10^{-4})$ & $\mathbf{0.5565}\;(1.5\!\times\!10^{-4})$ \\
 & Statin & $0.5354\;(4.3\!\times\!10^{-4})$ & $0.3318\;(6.6\!\times\!10^{-4})$ & $0.3501\;(3.4\!\times\!10^{-3})$ & $0.3024\;(7.1\!\times\!10^{-4})$ \\
\bottomrule

\end{tabular}
\endgroup
\caption{Macro F1 across prediction horizons on RIS\@. Cells report mean (sample variance; $n=10$). Bold indicates the highest mean within each setting and horizon.}
\label{tab:macro_f1_adni}
\end{table*}

\begin{table*}[p]
\centering
\begingroup
\fontsize{9pt}{9pt}\selectfont
\setlength{\tabcolsep}{3pt}

\begin{tabular}{llcccc}
\toprule
Model & Setting & 1-step & 2-step & 3-step & 4-step \\
\midrule
CRN & Test & $0.4254\;(7.9\!\times\!10^{-3})$ & $0.4321\;(1.4\!\times\!10^{-2})$ & $0.1500\;(8.4\!\times\!10^{-4})$ & $0.0000\;(0)$ \\
 & Placebo & $0.3954\;(1.0\!\times\!10^{-2})$ & $0.3791\;(8.4\!\times\!10^{-3})$ & $0.0120\;(3.7\!\times\!10^{-4})$ & $0.0031\;(9.5\!\times\!10^{-5})$ \\
 & Statin & $0.1993\;(2.7\!\times\!10^{-2})$ & $0.0250\;(1.0\!\times\!10^{-3})$ & $0.0140\;(1.2\!\times\!10^{-3})$ & $0.0000\;(0)$ \\
\addlinespace
GTrans & Test & $0.6358\;(7.4\!\times\!10^{-4})$ & $0.6607\;(6.0\!\times\!10^{-2})$ & $0.0900\;(2.5\!\times\!10^{-2})$ & $0.0436\;(1.5\!\times\!10^{-2})$ \\
 & Placebo & $0.6472\;(1.2\!\times\!10^{-3})$ & $0.5581\;(4.4\!\times\!10^{-2})$ & $0.0900\;(3.4\!\times\!10^{-3})$ & $0.0494\;(2.1\!\times\!10^{-2})$ \\
 & Statin & $\mathbf{0.4998}\;(1.4\!\times\!10^{-3})$ & $0.1910\;(2.0\!\times\!10^{-2})$ & $0.0300\;(1.5\!\times\!10^{-4})$ & $0.0000\;(0)$ \\
\addlinespace
TECDE & Test & $0.0533\;(1.6\!\times\!10^{-3})$ & $0.0100\;(5.01\!\times\!10^{-4})$ & $0.0227\;(8.6\!\times\!10^{-4})$ & $0.0096\;(4.1\!\times\!10^{-4})$ \\
 & Placebo & $0.0563\;(1.1\!\times\!10^{-3})$ & $0.0089\;(3.1\!\times\!10^{-4})$ & $0.0000\;(0)$ & $0.0000\;(0)$ \\
 & Statin & $0.1885\;(1.7\!\times\!10^{-2})$ & $0.0377\;(1.1\!\times\!10^{-3})$ & $0.0229\;(2.3\!\times\!10^{-3})$ & $0.0222\;(4.9\!\times\!10^{-3})$ \\
\addlinespace
RMSN & Test & $0.5759\;(2.9\!\times\!10^{-4})$ & $0.7508\;(1.2\!\times\!10^{-3})$ & $0.1127\;(2.6\!\times\!10^{-3})$ & $0.0430\;(2.3\!\times\!10^{-4})$ \\
 & Placebo & $0.5278\;(3.5\!\times\!10^{-4})$ & $0.6171\;(2.4\!\times\!10^{-3})$ & $0.0707\;(2.6\!\times\!10^{-4})$ & $0.0859\;(7.5\!\times\!10^{-4})$ \\
 & Statin & $0.2915\;(2.3\!\times\!10^{-3})$ & $0.2131\;(1.3\!\times\!10^{-3})$ & $0.1118\;(4.3\!\times\!10^{-6})$ & $0.0000\;(0)$ \\
\addlinespace
GNET & Test & $0.6328\;(5.3\!\times\!10^{-4})$ & $0.7160\;(2.9\!\times\!10^{-4})$ & $0.2700\;(5.1\!\times\!10^{-3})$ & $0.4233\;(5.0\!\times\!10^{-3})$ \\
 & Placebo & $0.6687\;(9.6\!\times\!10^{-4})$ & $0.6478\;(7.2\!\times\!10^{-4})$ & $0.3057\;(1.5\!\times\!10^{-3})$ & $0.3345\;(4.0\!\times\!10^{-3})$ \\
 & Statin & $0.4487\;(1.4\!\times\!10^{-3})$ & $\mathbf{0.2184}\;(2.3\!\times\!10^{-3})$ & $0.1572\;(1.0\!\times\!10^{-2})$ & $0.0250\;(6.3\!\times\!10^{-3})$ \\
\addlinespace
CHDT & Test & $\mathbf{0.7380}\;(1.3\!\times\!10^{-4})$ & $\mathbf{0.7518}\;(1.1\!\times\!10^{-3})$ & $\mathbf{0.5346}\;(1.1\!\times\!10^{-3})$ & $\mathbf{0.6458}\;(8.9\!\times\!10^{-4})$ \\
 & Placebo & $\mathbf{0.7384}\;(2.1\!\times\!10^{-4})$ & $\mathbf{0.6481}\;(9.4\!\times\!10^{-5})$ & $\mathbf{0.4564}\;(1.8\!\times\!10^{-3})$ & $\mathbf{0.5337}\;(3.8\!\times\!10^{-4})$ \\
 & Statin & $0.4491\;(1.7\!\times\!10^{-3})$ & $0.1656\;(1.5\!\times\!10^{-3})$ & $\mathbf{0.2343}\;(2.1\!\times\!10^{-2})$ & $\mathbf{0.1944}\;(1.1\!\times\!10^{-2})$ \\
\bottomrule

\end{tabular}
\endgroup
\caption{Dementia F1 across prediction horizons on RIS\@. Cells report mean (sample variance; $n=10$). Bold indicates the highest mean within each setting and horizon.}
\label{tab:dementia_f1_adni}
\end{table*}

\begin{table*}[p]
\centering
\begingroup
\fontsize{9pt}{11pt}\selectfont
\setlength{\tabcolsep}{3pt}

\begin{tabular}{llcccc}
\toprule
Model & Setting & 1-step & 2-step & 3-step & 4-step \\
\midrule
CRN & Test & $5.0\!\times\!10^{-9}$ & $1.8\!\times\!10^{-6}$ & $1.2\!\times\!10^{-18}$ & $7.4\!\times\!10^{-18}$ \\
 & Placebo & $1.4\!\times\!10^{-9}$ & $4.8\!\times\!10^{-5}$ & $6.8\!\times\!10^{-17}$ & $6.2\!\times\!10^{-17}$ \\
 & Statin & $1.2\!\times\!10^{-3}$ & $3.8\!\times\!10^{-5}$ & $0.24$ & $0.30$ \\
\addlinespace
GTrans & Test & $4.0\!\times\!10^{-10}$ & $0.19$ & $1.0\!\times\!10^{-17}$ & $6.6\!\times\!10^{-10}$ \\
 & Placebo & $6.1\!\times\!10^{-9}$ & $0.26$ & $3.5\!\times\!10^{-15}$ & $1.1\!\times\!10^{-7}$ \\
 & Statin & $3.5\!\times\!10^{-5}$ & $1.1\!\times\!10^{-5}$ & $0.16$ & $1.7\!\times\!10^{-5}$ \\
\addlinespace
TECDE & Test & $3.2\!\times\!10^{-11}$ & $3.7\!\times\!10^{-18}$ & $1.1\!\times\!10^{-18}$ & $2.7\!\times\!10^{-18}$ \\
 & Placebo & $8.9\!\times\!10^{-14}$ & $3.0\!\times\!10^{-18}$ & $1.1\!\times\!10^{-15}$ & $7.5\!\times\!10^{-21}$ \\
 & Statin & $0.414$ & $9.3\!\times\!10^{-5}$ & $0.38$ & $0.01$ \\
\addlinespace
RMSN & Test & $1.3\!\times\!10^{-19}$ & $0.26$ & $2.9\!\times\!10^{-15}$ & $1.8\!\times\!10^{-17}$ \\
 & Placebo & $2.1\!\times\!10^{-19}$ & $0.02$ & $4.8\!\times\!10^{-12}$ & $1.2\!\times\!10^{-17}$ \\
 & Statin & $0.39$ & $1.3\!\times\!10^{-7}$ & $6.4\!\times\!10^{-3}$ & $4.1\!\times\!10^{-9}$ \\
\addlinespace
GNET & Test & $4.1\!\times\!10^{-11}$ & $1.2\!\times\!10^{-8}$ & $5.0\!\times\!10^{-10}$ & $3.1\!\times\!10^{-9}$ \\
 & Placebo & $3.7\!\times\!10^{-11}$ & $0.18$ & $6.5\!\times\!10^{-11}$ & $5.4\!\times\!10^{-14}$ \\
 & Statin & $2.1\!\times\!10^{-4}$ & $1.8\!\times\!10^{-3}$ & $0.24$ & $0.29$ \\
\bottomrule
\end{tabular}
\endgroup
\caption{Exploratory two-sided Welch $t$-test $p$-values for Macro F1 versus CHDT on RIS ($n=10$ runs per model). Numerical $p$-values are reported.}
\label{tab:macro_f1_pvalues_adni}
\end{table*}

\begin{table*}[p]
\centering
\begingroup
\fontsize{9pt}{11pt}\selectfont
\setlength{\tabcolsep}{3pt}

\begin{tabular}{llcccc}
\toprule
Model & Setting & 1-step & 2-step & 3-step & 4-step \\
\midrule
CRN & Test & $1.2\!\times\!10^{-6}$ & $5.8\!\times\!10^{-6}$ & $5.4\!\times\!10^{-16}$ & $1.5\!\times\!10^{-13}$ \\
 & Placebo & $1.4\!\times\!10^{-6}$ & $5.9\!\times\!10^{-6}$ & $3.3\!\times\!10^{-13}$ & $5.9\!\times\!10^{-19}$ \\
 & Statin & $8.5\!\times\!10^{-4}$ & $8.9\!\times\!10^{-8}$ & $6.6\!\times\!10^{-4}$ & $2.1\!\times\!10^{-4}$ \\
\addlinespace
GTrans & Test & $1.4\!\times\!10^{-7}$ & $0.27$ & $6.4\!\times\!10^{-6}$ & $2.8\!\times\!10^{-8}$ \\
 & Placebo & $5.6\!\times\!10^{-6}$ & $0.21$ & $7.3\!\times\!10^{-11}$ & $1.9\!\times\!10^{-6}$ \\
 & Statin & $9.2\!\times\!10^{-3}$ & $0.60$ & $6.6\!\times\!10^{-4}$ & $2.1\!\times\!10^{-4}$ \\
\addlinespace
TECDE & Test & $6.2\!\times\!10^{-14}$ & $1.2\!\times\!10^{-13}$ & $4.7\!\times\!10^{-18}$ & $1.3\!\times\!10^{-19}$ \\
 & Placebo & $1.6\!\times\!10^{-16}$ & $2.0\!\times\!10^{-21}$ & $7.3\!\times\!10^{-11}$ & $1.8\!\times\!10^{-14}$ \\
 & Statin & $9.2\!\times\!10^{-5}$ & $3.4\!\times\!10^{-7}$ & $6.7\!\times\!10^{-20}$ & $4.9\!\times\!10^{-4}$ \\
\addlinespace
RMSN & Test & $5.0\!\times\!10^{-14}$ & $0.95$ & $4.7\!\times\!10^{-13}$ & $2.7\!\times\!10^{-17}$ \\
 & Placebo & $1.1\!\times\!10^{-15}$ & $0.08$ & $7.2\!\times\!10^{-12}$ & $5.1\!\times\!10^{-18}$ \\
 & Statin & $3.2\!\times\!10^{-7}$ & $0.01$ & $0.03$ & $2.1\!\times\!10^{-4}$ \\
\addlinespace
GNET & Test & $8.0\!\times\!10^{-9}$ & $0.01$ & $1.0\!\times\!10^{-7}$ & $8.4\!\times\!10^{-7}$ \\
 & Placebo & $2.4\!\times\!10^{-5}$ & $0.97$ & $1.3\!\times\!10^{-7}$ & $1.5\!\times\!10^{-6}$ \\
 & Statin & $0.98$ & $0.02$ & $0.19$ & $7.2\!\times\!10^{-4}$ \\
\bottomrule
\end{tabular}
\endgroup
\caption{Exploratory two-sided Welch $t$-test $p$-values for Dementia F1 versus CHDT on RIS ($n=10$ runs per model). Numerical $p$-values are reported.}
\label{tab:dementia_f1_pvalues_adni}
\end{table*}

\end{document}